\documentclass[prd,preprint,superscriptaddress,showpacs]{revtex4}
\usepackage{graphicx}
\newcommand{\be}{\begin{equation}}
\newcommand{\ee}{\end{equation}}

\begin{document}

\title{Notes on  $f(T)$ Theories}
\author{Yi Zhang}
 \email{zhangyia@cqupt.edu.cn}
 \affiliation{Department of Astronomy, Beijing Normal university,  Beijing 100875, China}
\affiliation{College of Mathematics and Physics, Chongqing University of Posts and Telecommunications,
Chongqing 400065, China}
\author{Hui Li}
\email{lihui@ytu.edu.cn}
\affiliation{Department of Physics, Yantai University, Yantai 264005,
China}
\author{Yungui Gong}
\email{gongyg@cqupt.edu.cn}
 \affiliation{College of Mathematics and Physics, Chongqing University of Posts and Telecommunications,
Chongqing 400065, China}
\author{Zong-Hong
Zhu}
\email{zhuzh@bnu.edu.cn}
\affiliation{Department of Astronomy, Beijing Normal university,  Beijing 100875, China}

\begin{abstract}

The cosmological models based on teleparallel
gravity with  nonzero torsion are considered.
 To investigate the evolution of this theory, we consider the phase-space analysis of the $f(T)$ theory.  It shows when the tension scalar can be written as an   inverse function of $x$ where $x=\rho_{e}/(3m_{pl}^{2}H^{2})$ and $T=g(x)$, the system is an autonomous one.
 Furthermore,
 the $\omega_{e}-\omega'_{e}$ phase analysis  is given out.
We perform the dynamical analysis for the models $f(T)=\beta T\ln(T/T_{0})$  and $f(T)=\alpha
m_{pl}^{2}(-T/m_{pl}^{2})^{n}$ particularly. We find that the universe will settle into de-Sitter phase for
both models. And we have examined the evolution behavior of the power law form in the $\omega_{ep}-\omega'_{ep}$ plane.

\end{abstract}


\maketitle

\section{Introduction}\label{sec1}
Recent cosmic acceleration in our universe is
suggested by a combination of different cosmic probes that primarily
involves Supernova data \cite{Riess:1998cb,Perlmutter:1998np}. Besides the most well-known dark energy scenario
(e.g., the scalar field quintessence  model  \cite{quintessence})  as the mechanism  for late time cosmic acceleration, another  kind of  model based on infra-red modifications to general relativity (such as the $f(R)$ theory \cite{Nojiri:2006ri,Sotiriou:2008rp,Faraoni:2008mf,Schmidt:2006jt}) was also proposed.
The validity of general relativity on large
astrophysical or cosmological scale has never been tested but only
assumed, the recent cosmic acceleration in our universe might be
nothing but the signal of  breakdown of General Relativity at large
scale
\cite{Briscese:2007cd,Hu:2007nk,Starobinsky:2007hu,Appleby:2007vb}.

The Riemann-Cartan geometry was proposed with the aim of  unifying gravity
and electromagnetism \cite{Einstein}.
In general,  a large number of connections can be defined  on a manifold. The assumptions of torsion-free and metric
 compatibility lead to the Einstein general relativity with Levi-Civita connection. Meanwhile, the assumption of curvature-free leads to  the teleparallel theory of gravity with Weitzenb\"{o}ck connection \cite{Hehl:1976kj,Hayashi:1979qx,Flanagan:2007dc,Garecki:2010jj}.
The $f(T)$ theory is proved to  be equivalent to general relativity but with different origins.
The $f(T)$ theory is considered as a gauge theory and general relativity is thought as a geometric theory.
Recently, the $f(T)$ gravity has been proposed to explain the present
 cosmic accelerating expansion without dark energy  \cite{inflation,Bengochea:2010sg,Wu,Yerzhanov,Linder,Bamba:2010wb}. Similar to $f(R)$ gravity, based on a modification of the teleparallel equivalent of general
relativity Lagrangian, the additional term related to the form of $f(T)$ could be written as  an effective
energy.
 The torsion $T$ in $f(T)$ theory will be  responsible for
the observed acceleration of the universe.

The $f(T)$ theory was firstly used as the source of driving inflation \cite{inflation}.
Then it was applied to  the late time acceleration \cite{Bengochea:2010sg,Linder}. Furthermore,
 the perturbation of $f(T)$ theory was discussed in Refs. \cite{Dent:2010va,Zheng:2010am},
the reconstruction of the $f(T)$ theory was presented in Refs. \cite{Karami:2010bu,Karami:2010xy},
and the observational constraint was  analyzed in Refs. \cite{Li:2010cg,Yang}.
In this letter, we made a dynamical analysis on two $f(T)$ models
by choosing variables different from those in Refs. \cite{Wu,dynamics}.

This letter is organized  as follows.  In
Sec. \ref{sec2}, we introduce the $f(T)$ theory, present the phase-space analysis and give out the analytical form of  $\omega'_{e}$.
Two particular models are discussed in
Sec. \ref{sec5}. Finally, we draw our conclusion in Sec. \ref{sec6}.

\section{The $f(T)$ Gravity}\label{sec2}
 In $f(T)$ theory,
the fundamental dynamical object is the vierbein field $e_{\mu}^{A}(x)$. In the teleparallelism, the orthonormal tetrad components $e_A^{\mu}(x)$ are used, where the index $A$ runs over 0, 1, 2, 3 in
the tangent space at each point $x^{\mu}$ of the manifold,
and $\mu$  is the coordinate index in the manifold and
also runs over 0, 1, 2, 3.
The curvature tensor and the covariant derivatives of $e_{A}^{\mu}(x)$ with
respect to the connection vanish globally, therefore $e_{A}^{\mu}(x)$ are absolutely parallel
vector fields, this theory is also called teleparallel gravity \cite{Einstein} and the geometry
is the Weitzenb\"{o}ck space-time characterized by  torsion tensor alone.
The vierbein field is related with the metric $g_{\mu\nu}$ by
\begin{equation}
\label{metric1}
g_{\mu\nu}=\eta_{A B} e^A_\mu e^B_\nu.
\end{equation}
The torsion $T^\rho_{\verb| |\mu\nu}$ and contorsion
$K^{\mu\nu}_{\verb|  |\rho}$ tensors are defined as
\begin{eqnarray}
&& T^\rho_{\verb| |\mu\nu} = e^\rho_A \left( \partial_\mu e^A_\nu -
\partial_\nu e^A_\mu \right), \\
&& K^{\mu\nu}_{\verb|  |\rho}=-\frac{1}{2} \left(T^{\mu\nu}_{\verb|  |\rho} -
T^{\nu \mu}_{\verb|  |\rho} - T^{\mu \nu}_{\verb|  |\rho}\right)\,.
\end{eqnarray}
The action for $f(T)$ theory is
\begin{equation}
I_{T}=\frac{1}{16 \pi G}\int d^4x |e| \left( T + f(T) \right)\,,
\end{equation}
where $|e|= \det \left(e^A_\mu \right)=\sqrt{-g}$,
the torsion scalar is
\begin{equation}
\label{}
 T \equiv S^{\mu\nu}_{\verb|  | \rho} T^{\rho}_{\verb|  |
\mu\nu}\,,
\end{equation}
and
\begin{equation}
\nonumber S^{ \mu\nu}_{\verb|  |\rho}\equiv \frac{1}{2} \left(K^{\mu\nu}_{\verb|  |\rho}
+\delta^\mu_\rho \ T^{\alpha \nu}_{\verb|  | \alpha} - \delta ^\nu _
\rho \ T ^{\alpha \mu }_{\verb|  | \alpha }\right)\,.
\end{equation}

Varying the action with respect to the vierbein yields the equation of motion,
 \begin{eqnarray}
 \label{me}
 \nonumber
 &&-\frac{1}{4}e^{A}_{\verb| |\mu}(T+f)+e^{\beta}_{\verb| |A}T^{\mu}_{\verb| |\nu\beta}S_{\mu}^{\verb| |\nu\alpha}
 (1+f_{T})\\
 &&+e^{-1}\partial_{\mu}(ee^{\rho}_{A}S_{\verb|  |\rho}^{\mu \alpha})(1+f_{T})+e^{\rho}_{A}S_{\rho}^{\verb| |\mu\alpha}
 f_{TT}\partial_{\mu}T=4\pi G e^{\rho}_{\verb| |A}T_{\rho}^{\verb| |\alpha}
 \end{eqnarray}
 where $T_{\rho}^{\alpha}$
is the energy-momentum tensor and the subscript $T$ denotes the derivative with respect to the torsion scalar.

For simplicity, we assume a flat Friedmann-Robertson-Walker metric,
\begin{equation}
{ds}^2 = -{dt}^2 + a^{2}(t)\sum^{3}_{i=1}(dx^{i})^{2}\,,
\end{equation}
where $a(t)$ is the scale factor. In the FRW space-time, $g_{\mu \nu}= \mathrm{diag} (-1, a^2, a^2,a^2)$
and therefore the tetrad components $e^A_\mu = (1,a,a,a)$ yield the
torsion scalar $T=-6H^2$, where $H=\dot{a}/a$ is the
Hubble parameter.

The energy conservation equations for the radiation and  pressureless matter are
\begin{eqnarray}
 \label{rgprime}
 &&\rho_{\gamma}'+4\rho_{\gamma}=0,\\
 \label{rmprime}
 &&\rho_{m}'+3\rho_{m}=0,
 \end{eqnarray}
 where $\rho_{\gamma}$ and $\rho_{m}$ are the energy densities of radiation  and pressureless matter respectively,
and a prime means the derivative with respect to $\ln a$.
From Eq. (\ref{me}),  the Friedmann equation for $f(T)$ theory is gotten,
 \be
 \label{H2}
H^{2}=\frac{8\pi G}{3}(\rho_{m}+\rho_r+\rho_{e}),
 \ee
where the effective (dark) energy density due to $f(T)$ is
   \be
   \rho_{e}=\frac{1}{16\pi G}(-f+2Tf_{T}).
   \ee
Obviously,  when $f=2Tf_{T}$, i.e., $f(T)\propto T^{1/2}$,  the effective energy density vanishes.
The effective dark energy density is assumed to be conserved as well
 \be
 \label{rhoec}
\rho_{e}'+3(\rho_{e}+p_{e})=0,
 \ee
where the effective pressure is $ p_{e}=-\rho_{e}-m_{pl}^{2}(f_{T}+2Tf_{TT})T'/6$, and
the evolution of the torsion scalar $T$ could be expressed as
  \be
  \label{Tprime}
 \frac{T'}{T}=\frac{-1}{3m_{pl}^{2}H^{2}}\frac{3\rho_{m}+4\rho_r}{2Tf_{TT}+f_{T}+1}.
 \ee

\subsection{The Phase-Space Analysis}\label{sec3}

To make the phase-space analysis, we introduce the following dimensionless variable,
 \begin{eqnarray}
 \label{xyz}
  x=\frac{\rho_{e}}{3m_{pl}^{2}H^{2}}, \quad y=\Omega_m=\frac{\rho_{m}}{3m_{pl}^{2}H^{2}},\quad z=\Omega_r=\frac{\rho_r}{3m_{pl}^{2}H^{2}}.
 \end{eqnarray}
In terms of these dimensionless variables, the Friedmann equation becomes the constraint equation
 \be
 \label{Friedmann}
 x+y+z=1.
 \ee

Combining Eqs. (\ref{rgprime}), (\ref{rmprime}), (\ref{rhoec}) and (\ref{Tprime}), we get
  \begin{eqnarray}
  \label{xprime}
  x'&=&(f_{T}-\frac{f}{T}-2Tf_{TT})\frac{T'}{T},\\
 \label{yprime}
  y'&=&-y(3+\frac{T'}{T}),\\
\label{zprime}
 z'&=&-z(4+\frac{T'}{T}),
 \end{eqnarray}
and
 \be
 \label{Hprime}
 \frac{T'}{T}= \frac{(H^{2})'}{H^{2}}=-\frac{3y+4z}{2Tf_{TT}+1+f_{T}}.
  \ee
From the definition of $x$ in Eq.(\ref{xyz}), we know that once the function $f(T)$ is specified, $x$ can be
expressed as a function of $T$ , so if  the inverse function exists, then $T$ can be expressed as a function of $x$,
 \be
 \label{inverse}
  T=g(x).
   \ee
With the help of the above relation, Eqs. ({\ref{xprime}), (\ref{yprime}) and (\ref{zprime})
become  an autonomous system.
Because of the constraint (\ref{Friedmann}), there are only two independent variables, here we choose them
as $x$ and $y$.

\subsection{The Equation of State}
The elucidation of the physics of models becomes difficult when the EoS parameter is close to $-1$ \cite{Riess:2004nr},
since the evolving slowly effective energy density  may be indistinguishable from  $\Lambda$CDM scenario.
By emphasizing the dynamics, the restricted regions of the trajectories are recovered in ``position'' and ``velocity''-
the value of the equation of state ratio $\omega$ and its time variation $\omega'$ \cite{Caldwell:2005tm}.

The EoS parameter of the effective dark energy is,
 \be
 \label{we}
 w_{e}=\frac{p_{e}}{\rho_{e}}=-1+\frac{T'}{T}\frac{f_{T}+2Tf_{TT}}{3(-f/T+2f_{T})}.
 \ee
 When $f(T)=constant$, the model can be regarded as the cosmological constant theory which  is attributed  to the quantum zero-point
energy of the particle physics vacuum.

It is currently
constrained from the distance measurements of $SNIa$ as $\omega_{0}=-1.31^{+0.22}_{-0.28}$ and $\omega_{0}'=-1.48^{+0.90}_{-0.81}$ \cite{Riess:2004nr}.
In $f(T) $ theories,  the  form of $\omega_{e}'$ can be expressed as
 \be
 \label{omegaprime}
 \omega_{e}'=\frac{f_{T}+2Tf_{TT}}{3(-f/T+2f_{T})}(\frac{T'}{T})'+\left(\frac{3Tf_{TT}+
 2T^{2}f_{TTT}}{3(-f/T+2f_{T})}-\frac{Tf_{T}+2T^{2}f_{TT}}{3(-f/T+2f_{T})^{2}}\right)\frac{T'^{2}}{T}.
 \ee
It is noticed that the parameters  $\omega_{e}$ and $\omega_{e}'$ are  functions of $T$,
$\rho_{m}$ and $\rho_{\gamma}$. But  only the Friedmann equation  could
be used as a constraint equation.  It is hard to draw the phase-space diagram since both the parameters $\omega_{e}$ and $\omega_{e}'$ are
 functions of two variables. Fortunately,
 in the pivot periods, we could get the relation between
 $\omega_{e}-\omega_{e}'$ once the exact form of $f(T)$ is given. In the radiation dominated phase, the
 energy of matter could be ignored; and once the matter is dominating,
 the energy density of radiation could be ignored.
 In the following , we will give two exact forms of $f(T)$ as examples.

\section{Two models}\label{sec5}
\subsection{The Logarithmic Form}
Firstly, we consider a phenomenological model with a logarithmic form,
   \be
  f(T)=\beta T\ln(\frac{T}{T_{0}}),
   \ee
where $\beta$ and $T_{0}$ are  constants.
The effective energy density is
   \be
   \rho_{el}=\frac{1}{16\pi G}\left(\beta T\ln(\frac{T}{T_{0}})+2 \beta T\right),
   \ee
and the dimensional variable $x$  is
   \be
 x=\frac{2}{3}\beta \ln \frac{T}{T_{0}}-\frac{\beta}{3}.
 \ee
Therefore,
   \be
   T=T_{0}e^{\frac{3}{2\beta}x+\frac{1}{2}},
   \ee
and
  \be
  \frac{T'}{T}=-\frac{4-4x-y}{3\beta-1-x/\beta}.
  \ee
The dynamical equations of the autonomous system become
 \begin{eqnarray}
 \label{xl}
  x'&=&- \beta \frac{4-4x-y}{3\beta-1-x/\beta}, \\
  \label{yl}
 y'&=&-y(3-\frac{4-4x-y}{3\beta-1-x/\beta}).
  \end{eqnarray}
Setting $x'=y'=0$,  we get the critical point $x=1$ and $y=0$. The critical point corresponds to dark energy domination with $\Omega_m=\Omega_r=0$ and $w_e=-1$. The eigenvalues are $\lambda_1=-3$ and
$\lambda_2=4\beta^2/(3\beta^2-\beta-1)$, so when $(1-\sqrt{13})/6<\beta<(1+\sqrt{13})/6$ and $\beta\neq 0$,
the critical point is a stable fixed point. Unlike the standard model, there is no critical points for radiation domination $z=1$ and matter domination $y=1$ for this model. In the view of dynamical analysis, this model is ruled out by the history of our universe.
 Therefore, it is not necessary to discuss its EoS parameter.

\subsection{The Power Law Form}
Now we consider the power law form,
 \be
 f(T)=\alpha m_{pl}^{2}(\frac{-T}{m_{pl}^{2}})^{n},
   \ee
where $\alpha$ and $n$ are dimensionless parameters.
When $n=0$, $f(T)$ is a constant and the model becomes the $\Lambda$CDM model.
By using the power-law form of $f(T)$,  we get the effective dark energy density,
 \be
 \label{prho}
 \rho_{ep}=\alpha m_{pl}^{4}(n-\frac{1}{2})(\frac{-T}{m_{pl}^{2}})^{n},
 \ee
 and the effective pressure,
 \be
 \label{pep}
 p_{ep}=-\alpha m_{pl}^{4}(n-\frac{1}{2})(\frac{-T}{m_{pl}^{2}})^{n}+\frac{ (2n^{2}-n)\alpha m_{pl}^{2}}{6}(\frac{-T}{m_{pl}^{2}})^{n-1}T'.
 \ee
When $n=1/2$, $\rho_{ep}=p_{ep}=0$. When $n=1$,  $\rho_{ep}=-\alpha m_{pl}^2 T/2=3\alpha m_{pl}^2 H^2$,
so the effective energy tracks the background matter. Substituting Eq. (\ref{prho}) into the definition of the dimensionless
variable $x$, we get
 \be
 x=\alpha(2n-1)(\frac{-T}{m_{pl}^{2}})^{n-1},
 \ee
and
 \be
 T=-m_{pl}^{2}\left(\frac{x}{\alpha(2n-1)}\right)^{1/(n-1)}.
 \ee
So
 \be
 \frac{T'}{T}=-\frac{4-4x-y}{1-n x},
 \ee
and the dynamical equations are,
  \begin{eqnarray}
  \label{Px}
   x'&=&-\frac{2(n-1)(4-4x-y)}{1-n x}x,\\
   \label{Py}
   y'&=&-y(3-\frac{4-4x-y}{1-n x}).
   \end{eqnarray}
Setting $x'=y'=0$, we obtain the critical pints $(x,\ y)=(0,\ 0)$, $(x,\ y)=(0,\ 1)$ and $(x,\ y)=(1,\ 0)$.
The properties of the fixed points $R$, $M$, and $A$  are summarized in  Table \ref{power1}.
These fixed points are similar to  those found in Ref.\cite{Wu}, but  we find the fixed point
$M$ can be stable when $n^2>1$.

\begin{table}
\begin{center}
\begin{tabular}{|c|c|c|c|}
\hline
 Phases &$(x,y)$ & Stability &$(\lambda_{1},\lambda_{2})$\\
\hline
\hline  $R$&$(0,0)$  & unstable  & $(1,\ -8(n^{2}-1))$ \\
\hline  $M$&$(0,1)$ & unstable when $n^{2}<1$ & $(-1,\ 6(1-n^{2}))$ \\
\hline  $A$&$(1,0)$  & stable  & $(-3,\ -8)$  \\
\hline
\end{tabular}
\end{center}
\caption{The properties of the critical points for the power law form with $n\neq1$.}
\label{power1}
\end{table}

\begin{figure}
  \center
  \includegraphics[width=0.5\textwidth]{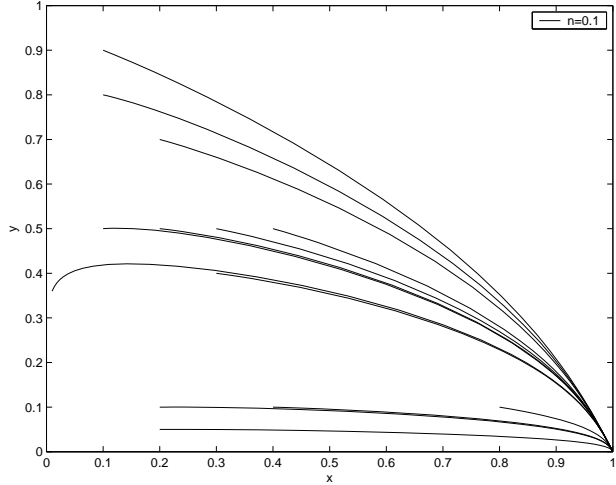}
    \caption{\label{fixedpoint}The phase diagram of attractor $A$ for power law $f(T)$ form with  different initial conditions.  }
\end{figure}

\begin{description}
\item[R~~~]The radiation dominated phase\\
For the critical point $R$, $x=y=0$ and $\Omega_r=1$. Because $\lambda_1=1>0$,
it is an unstable fixed point, so the universe could exit from the radiation dominating phase.
\item[M~~~]The matter dominated phase\\
For the critical  point $M$, $x=0$ and $y=1$. When $n^2>1$, it is a stable attractor, otherwise
it is an unstable fixed point. So when $n^2<1$, the universe could exit from the matter domination.
\item[A~~~]The accelerating phase\\
For the fixed point $A$, $n\neq 1$, $x=1$ and $y=0$ and it is a stable attractor independent of
the parameters $\alpha$ and $n$.
So $\Omega_m=\Omega_r=0$, $w_e=-1$ since $T'/T=0$, and
this phase recovers the de-Sitter phase. Now we have two attractors $M$ and $A$ when $n^2>1$.
For a given initial condition, which attractor will the universe choose in the phase space?
To answer this question, we go back to the dynamical Eqs. (\ref{Px}) and (\ref{Py}). Note that
there is a singularity in the system when $x=1/n$. So if we start with $x>1/n$, then the attractor
is $A$; otherwise if we start with $x<1/n$, the attractor is $M$. When $n^2<1$,
we have only one attractor $A$. We plot the phase diagram for the
attractor $A$ with $n=0.1$ in Fig. \ref{fixedpoint}.
 \end{description}

For the special case $n=1$, if $\alpha<1$, $x=\alpha$.  The constraint equation becomes $y+z=1-\alpha$, so there is
only one independent variable. And there are only two critical points $R$
with $\Omega_r=1-\alpha$ and $M$ with $\Omega_m=1-\alpha$.

\begin{figure}
  \center
  \includegraphics[width=0.5\textwidth]{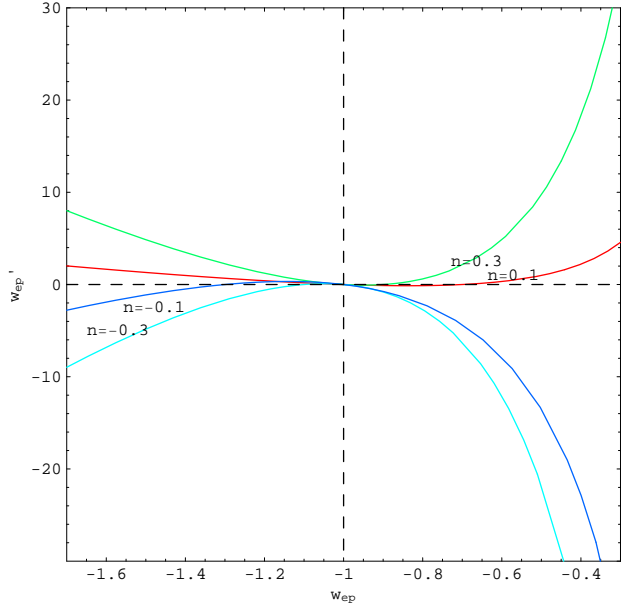}
    \caption{The trajectories of  $\omega_{ep}-\omega_{ep}'$  in the phase space. From top to bottom, $n$ takes the values of  0.3, 0.1, -0.1, -0.3, respectively.}
    \label{wwe}
\end{figure}

Based on the power law form of $f(T)$, the corresponding equation of state is obtained
    \begin{eqnarray}
    \label{w1}
\omega_{ep}=-1-\frac{n}{3}\frac{T'}{T}.
   \end{eqnarray}
In the radiation or the matter dominated phase, the evolution of the scale factor $a$ is known, then it is easily to get the effective EoS  parameter which is listed in Table \ref{tab3}.  Generally, if the effective energy density part takes the dominated part, the corresponding effective EoS parameter $\omega_{el}<0$ which suggests $n<1$.
If the effective energy density part has the possibility to make our universe accelerate, the corresponding effective EoS parameter $\omega_{ep}<-1/3$ which suggests $n<2/3$. If the effective EoS parameter $\omega_{ep}<-1$,  $n<0$ is suggested.

\begin{table*}[t]
\begin{center}
\begin{tabular}{|c|c|c|c|c|c|}
\hline
 Phases & Scale factor &$T'/T$& $\omega_{el}$ & $\omega_{el}'$\\
\hline
\hline  R&$a\propto t^{1/2}$ &$-4$ &$-1+4n/3$ & $0$\\
\hline  M &$a\propto t^{2/3}$ &$-3$ & $-1+n$ & $0$  \\
\hline
\end{tabular}
\end{center}
\caption[crit]{\label{tab3} we summarize  the critical points of the power law form.}
\end{table*}

After the universe  transits to the matter dominated phase,  the radiation part could be  ignored, and the energy density of matter could be approximately  written as
 \be
\rho_{m}\simeq \frac{m_{pl}^{2}}{2}(-T+f-2Tf_{T}).
 \ee
Putting the above equations into Eq.(\ref{we}),
 the equation of state is expressed as
 \begin{eqnarray}
 \nonumber
 \omega_{ep}\simeq \frac{n-1}{1-\alpha n(2n-1)(-T/m_{pl}^{2})^{n-1}}-1-\frac{n}{3}\frac{T'}{T}.
 \end{eqnarray}
 The derivative of the EoS parameter with respect to $\ln a$ is changed to
\begin{eqnarray}
\label{ep}
\omega_{ep}'\simeq\frac{3(n-1)^{2}}{n}(1+\omega_{ep})(\frac{n-1}{\omega_{ep}}-1).
\end{eqnarray}
After fixing the value of $n$,   the diagram of $\omega_{ep}-\omega_{ep}'$ which is Figure \ref{wwe} could be drawn out.  There is a fixed point $\omega_{ep}=-1,\omega_{ep}'=0$ which is corresponding to the accelerating phase $A$.

 \section{Conclusion}\label{sec6}

 In this letter,  the phase-phase analysis of the $f(T)$ theories  is done and its stability of the critical points has been investigated. We can see
 that when the energy density can be written as an   inverse function of $x$  which could be expressed as a general form
 $T=g(x)$, we can do the phase-analysis in the dynamical system. Furthermore,  the value of the EoS parameter and its evolution are considered. Those analyses  depend on the exact forms of $f(T)$ heavily.

Specially, we made the dynamical analysis for the logarithmic form and the power form of $f(T)$. For the logarithmic form $f(T)=\beta T\ln(T/T_{0})$, only one critical point which corresponds to de-Sitter phase exists. Unlike the standard cosmology with Einstein gravity, there is no critical point for radiation and matter dominations.
For the power law form $f(T)=\alpha
m_{pl}^{2}(-T/m_{pl}^{2})^{n}$, we find that two stable fixed points exist when $n^2>1$.
One fixed point corresponds to matter domination and the other fixed point corresponds to de-Sitter phase.
The reason for the existence of two stable fixed points is that the autonomous system has a singular point $x=1/n$.
If the system starts with $x>1/n$, then the system will settle into de-Sitter phase $A$, otherwise the system
will settle into matter domination $M$. When $n=1$, de-Sitter phase is absent and the effective energy tracks the
background matter. When $n^2<1$, only de-Sitter phase is the stable fixed point. And we have plotted  the diagram of  $\omega_{ep}-\omega'_{ep}$ plane for the power law form.
The fixed point $\omega_{ep}=-1,\omega_{ep}'=0$  is corresponding to the accelerating phase $A$.


\acknowledgments
We thank the helpful comments from  the anonymous referee.
This work was supported by  the Ministry of Science and Technology of
China national basic science Program (973 Project) under grant Nos.
2007CB815401 and 2010CB833004, the National Natural Science
Foundation of China  key project  under grant No. 10935013,
the National Natural Science Foundation of China under grant
Nos.11005164 and 11073005,  the Distinguished Young Scholar Grant 10825313,
CQ CSTC under grant Nos. 2009BA4050 and 2010BB0408,
and CQ MEC under grant No. KJTD201016. YZ was partially supported by China Postdoc Grant
No.20100470237.

\end{document}